\documentclass[useAMS,usenatbib,usegraphicx]{mn2e}

\def\la{\raise.5ex\hbox{$<$}\kern-.8em\lower 1mm\hbox{$\sim$}}
\def\ma{\raise.5ex\hbox{$>$}\kern-.8em\lower 1mm\hbox{$\sim$}}

\def\msol{M$_{\odot}$ }
\def\kms{$\rm km\, s^{-1}$}
\def\cm3{$\rm cm^{-3}$}
\def\Ts{$\rm T_{*}$}
\def\Vs{$\rm V_{s}$}
\def\n0{$\rm n_{0}$}
\def\B0{$\rm B_{0}$}

\def\Ne{$\rm N_{e}$}
\def\Te{$\rm T_{e}$}

\def\erg{$\rm erg\, cm^{-2}\, s^{-1}$}
\def\mum{$\mu$m}

\def\mum{$\mu$m}

\def\L12{L$_{12\mu m}$~}
\def\F12{F$_{12\mu m}$~}

\def\Hb{H$\beta$~}
\def\Ha{H$\alpha$~}
\def\Ly{Ly$\alpha$~}
\def\ff{{\it ff}}

\title{Shock fronts in the symbiotic system BI Crucis}

\author[M.Contini, R.Angeloni, P.Rafanelli]{M. Contini$^{1}$, R. Angeloni$^{2,1}$, and P. Rafanelli$^{2}$ \\
$^{1}$School of Physics and Astronomy, Tel-Aviv University, Tel-Aviv, 69978 Israel\\
$^{2}$Dipartimento di Astronomia, Universit$\grave{a}$ di Padova, Vicolo dell'Osservatorio 2, I-35122 Padova, Italy\\
}
\begin{document}

\date{Accepted . Received ; in original form }

\pagerange{\pageref{firstpage}--\pageref{lastpage}} \pubyear{2002}

\maketitle

\label{firstpage}

\begin{abstract}
We  investigate the symbiotic star BI Crucis through a comprehensive and self-consistent analysis of the spectra 
emitted  in three different epochs: 60's, 70's, and late 80's.
In particular, we would like to find out the physical conditions in the shocked nebula and in the 
dust shells, as well as their location within the symbiotic system, by exploiting both photometric and 
spectroscopic data from radio to UV.
We suggest a model which, on the basis of optical imaging, emission 
line ratios and spectral energy distribution profile, is able to account for collision of the winds, 
formation of lobes and jets by accretion onto the WD, as well as for the interaction of the blast wave from a past, 
unrecorded outburst with the ISM.
We have found that the spectra observed throughout the years show the marks of the different processes at 
work within BI Cru, perhaps signatures of a post-outburst evolution. We then call for new infrared 
and millimeter observations, potentially able to resolve the inner structure of the symbiotic nebula.

\end{abstract}

\begin{keywords}
binaries: symbiotic - stars: individual: BI Cru
\end{keywords}

   \maketitle

\section{Introduction}
\begin{figure}
\begin{center}
\includegraphics[width=0.45\textwidth]{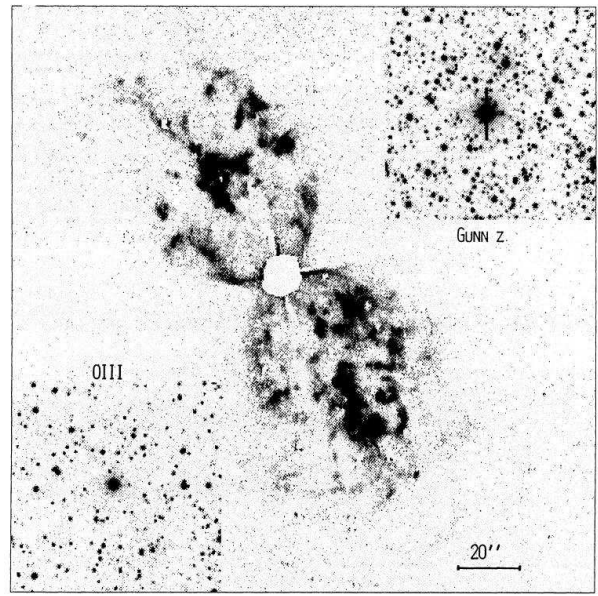}
\caption{The image of BI Cru, from SC92: central image is in [NII] with [NII] continuum
subtracted to remove the many background stars; top right is the Gunn z image,
and bottom left the [OIII] frame. \label{fig:CS92}}
\end{center}
\end{figure}

BI Crucis (BICru) is a dusty (D-type) symbiotic system (SS) (Kenyon et al. 1986)  which hosts an early Mira whose pulsation 
period is 280d (Whitelock et al. 1983),
and a hot star of \Ts $\sim$ 26500 K (Rossi et al. 1988, hereafter R88). 
With respect to other
dusty SSs, BI Cru shows a less strong IR excess which can be attributed
to thermal emission of relatively cool dust (Angeloni et al. 2007a).

The discovery of an associated bipolar nebula with a total extent of 1.3pc (Fig. \ref{fig:CS92})
by Schwarz \& Corradi (1992, hereafter SC92), pointed out a strong morphological similarity between BI Cru and He2-104, the Southern Crab.
However, the BI Cru nebula seems to have a dynamical age
of 3000 yrs, being thus at a slightly different evolutionary age with respect to the He2-104 one (Corradi \& Schwarz1993, hereafter CS93). 

Previous studies by Morris (1987) proposed a binary model for the formation of bipolar planetary nebulae via
variable accretion rates onto the WD. Jets and fast winds would be thus naturally created,
but for a meaningful modelling of BI Cru at least two other elements should be taken
into account, namely, the bipolar nebula expands at as high velocities as the
jets (200 \kms, CS93), and there are hints of multiple events,  such as periodic
(every $\sim$1000 yr, CS93) hydrogen shell flashes that may have occurred on the WD surface.

Bipolar jets and lobes suggest the presence
of an accretion disk, whose formation may be plausible assuming a typical accretion rate
of 10$^{-7}$ \msol yr$^{-1}$ (Morris 1987).
CS93 suggest that the fast winds from the hot star are produced by
thermonuclear runaways on the surface of the WD. Disk instabilities are less indicated
because stable hydrogen burning occurs only in a very little range of accretion rates, which would constrain
the binary parameters.

Previous modelling of SSs
in different phases of outburst and quiescence (Contini et al. 2009a, Angeloni et al 2007a, and references therein)
 led to
recognize some main dynamical mechanisms, that can be summarized by: the collision of the
stellar winds which leads to shocked nebulae at different location on the orbital plane,
the formation of a disk as a consequence of accretion phenomena, the ejection of jets and
lobes perpendicularly to the orbital plane, and the outburst of the WD, at the origin of the
blast wave propagation outwards in the ISM. Furthermore, also the dust shells emitted by
the Mira contribute to the line and continuum spectra and might be responsible for obscuration episodes.

In this paper, we investigate the origin of the emission fluxes at different epochs. 
analyzing the spectral and morphological appearance of BI Cru. 
On the basis of the observational and theoretical  evidences described previously, we will account 
for episodes of wind collisions, ejection of lobes and jets due to the
accretion processes, and expansion of the blast wave in the surrounding medium
as a consequence of past outbursts of the WD.

Quantitative informations  can be derived  only by modelling 
the 1962 spectrum presented  by Henize \& Carlson (1980, hereafter  HC80), which provides intensities and velocities of several observed lines.
Further, important informations  can be obtained by the observation of the broad \Ha ~
line reported by Whitelock et al. (1983) and by the polarization of its wings discussed by Harries (1996).
Eventually, some upper and/or lower limits to the physical parameters derived from the
spatial distribution of some important emission lines (e.g. [OII], [OIII]) were found  by SC92.

We adopt the models presented for He2-104 and for R Aqr by Contini \& Formiggini (2001 and 2003, respectively). There, the winds
from the WD and the red giant star collide head-on between the stars and head-on-back
outward the binary system, leading to a network of shock fronts in the equatorial plane of
the binary system (Girard \& Willson 1987).
Moreover, the jets from the accretion disk,  colliding with the circumstellar
matter,  give origin to the bipolar nebula in  the perpendicular direction (Contini \& Formiggini 2003). 
 In fact, the jet velocity and the velocities observed in the
bipolar nebula are similar ($\sim$ 200-250 \kms).

The modelling of line and continuum
spectra  makes use of SUMA\footnote{http://wise-obs.tau.ac.il/$\sim$marcel/suma/index.htm}, 
a code that simulates the physical conditions of an emitting gaseous
nebula under the coupled effect of photoionization from an external source and
shocks. The important role of dust is investigated following Angeloni et al. (2007a,b,c).

The observations of BI Cru at different epochs are presented in Sect. 2.
 The  1962 spectra   are analyzed in Sect. 3,  the broad \Ha line is extensively discussed in Sect. 4,
and the bipolar lobes are modelled in Sect. 5. The continuum
 spectral energy distribution (SED), calculated consistently with the line spectra,
 is compared with the data in Sect. 6. Discussion and concluding remarks follow in Sect. 7.

\section{Observational data}
SSs are rarely observed with a clear long-term strategy through the years. For  most
objects, the data from the literature are either the result of specific observations or
belong to large surveys of those generally said "peculiar emission-line stars" such
as PNe, Novae, CVs, etc. Unfortunately, this is also the case of BI Cru, since its
discovery on Michigan-Mount Wilson Southern \Ha Survey plates in 1949. In the following,
we summarize the data we have exploited in order to constrain our physical interpretation of this SS.

\begin{figure}[!hb]
\begin{center}
\includegraphics[width=0.45\textwidth]{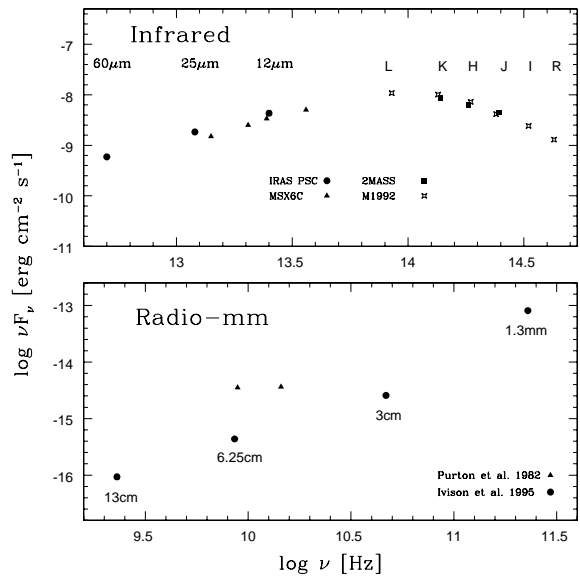}
\caption{Observational data. Top panel: IR spectral range. The 60 \mum\ IRAS point is an upper limit.  Bottom panel: radio spectral range.
The two data points from Purton et al. (1982, triangles) are also upper limits.\label{fig:varbi}}
\end{center}
\end{figure}

\subsection{Photometric data}
Besides the two upper limits (namely, missed detections) reported in Purton et al. (1982),
the only information we have about the radio-mm wavelength range in BI Cru comes from the
survey by Ivison et al. (1995 - Fig. \ref{fig:varbi}, bottom panel).
Unfortunately, this implies that there is no way to investigate any radio variability,  which may be
expected according to its specific nature of SS.

With regards to the IR photometric data, it is worth mentioning that there is a
reasonable agreement between observations taken at different times (Fig. \ref{fig:varbi}, top panel).
This means that our discussion of the continuum SED is not invalidated
by the well known IR variability mainly ascribed to the Mira pulsations.
The data come from the IRAS Point Source Catalog (the 60 \mum ~
point being actually an upper limit) and the Midcourse Space Experiment Point Source Catalog
(Version 2.3, Egan et al. 2003) for the longer wavelengths, and from Munari et al. (1992) and the 2MASS
survey for the NIR range.

\subsection{Spectroscopic data}
The  absence of a long-term strategy in observing BI Cru has not prevented to recognize an
intrinsic variability of both emission and absorption lines since the late '40s.
In particular, variations in \Ha\ intensity were suspected already by 
 HC80, which pointed out that the line  was stronger in 1950-51 than in 1949.
Moreover, they reported on a Mount Stromlo slit spectrogram obtained in 1962  which
shows an emission line spectrum of relatively high excitation, superposed on a weak bluish
continuum (HC80, Table 1).\\

In the 70's, Lee (1973) noticed on a spectrum taken in 1968 the presence of strong unusual
emission features, probably due to FeII, as well as a very strong \Hb, and a moderately
strong H$\gamma$. Allen (1974), on the other hand, recorded more than 40 lines of FeII,
[FeII], weak lines of HeI, and also suspected [OI], in addition to Balmer emission lines.
Interestingly, he found rather broad emission lines and a weak violet component to H$\gamma$
displaced from the principal line by several hundred \kms.

In any case, the most remarkable difference between the 1962 and 1974 spectra is the
presence of [OIII], [NeIII], and [SII] with different FWHM in 1962, and their absence in 1974 (HC80).

In the 80's, Whitelock et al. (1983) presented some infrared photometry, from which they
deduced the evidence of a possible secular decrease in intensity between 1979 and 1982,
as well as new optical spectra taken in 1974 just 55 days before the Allen's spectrum.
In these spectra, surprisingly,  a strong \Ha emission was evident  showing blue displaced P Cyg
absorption and extensive emission wings with FWHM of $\sim$ 1500-2000 \kms, and FeII lines
with a P Cyg profile where the E-A radial velocity difference was of 145 \kms.

\begin{figure}
\begin{center}
\includegraphics[width=0.45\textwidth]{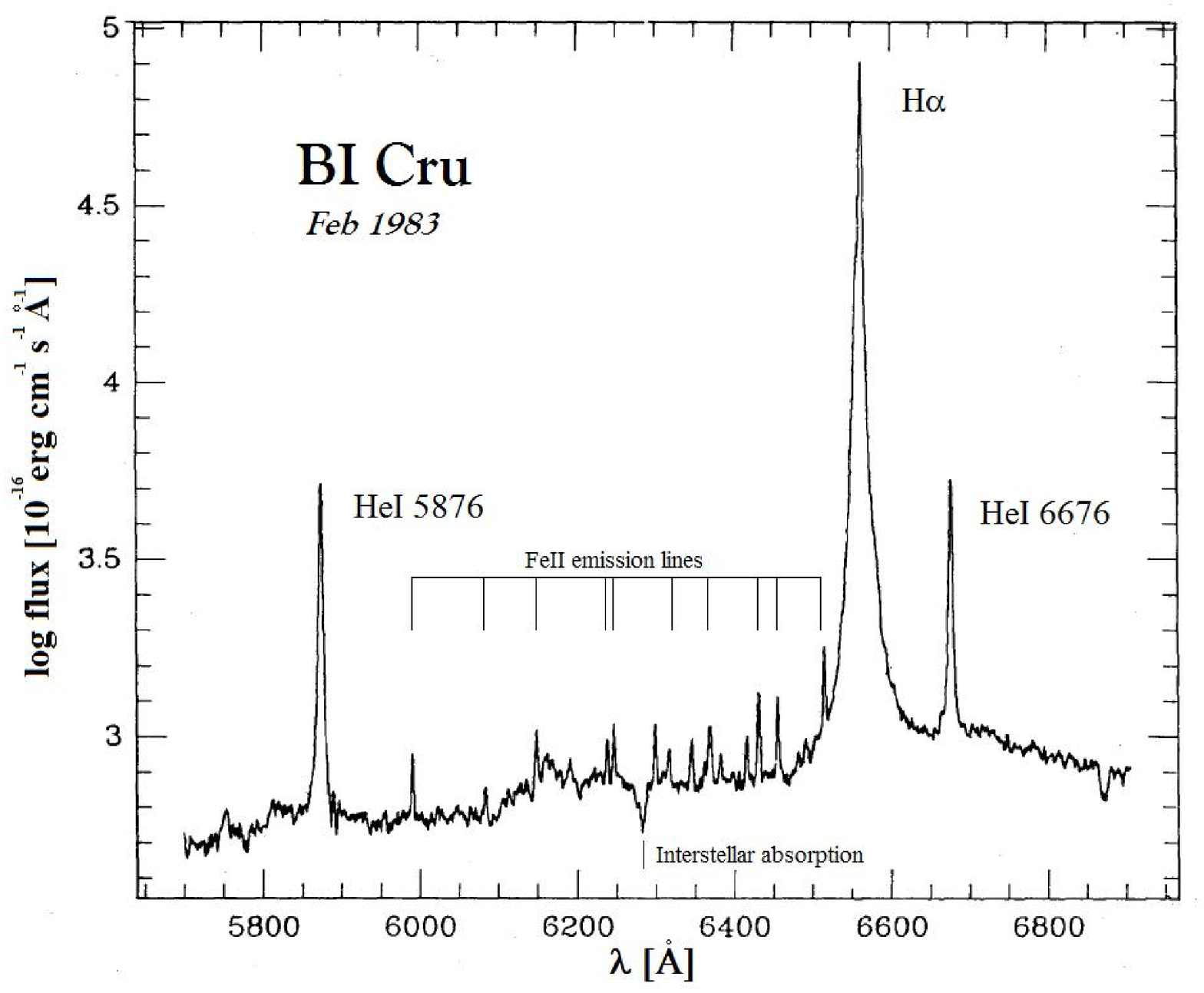}
\caption{Top: Low-resolution spectrum of BI Cru in Feb. 1983: \Ha and HeI 5876 and 6676 are very strong.
Several FeII emission lines are also noticeable, as well as the 6284 band of interstellar origin
(adapted from R88).\label{fig:rossi}}
\end{center}
\end{figure}

A few years later, R88 analyzed a spectrum taken in 1983 with the 1.5m ESO
telescope at La Silla. Strong \Ha\ emission and prominent HeI lines confirmed that BI Cru was
in a fairly high ionization stage.     Several FeII emission lines were also present (Fig. \ref{fig:rossi}).
The spectrum observed by R88 shows strong \Ha\ and blend of FeII.
R88 calculated a minimum black body (bb) T=26500 K for the hot star and referred
to two emission components in \Ha of 110 and 350 \kms, with a violet shifted absorption
extending from 0 to -300 \kms.

In the 90's, SC92 finally reported the discovery of a bipolar nebula
associated with BI Cru (Fig. \ref{fig:CS92}), whose morphology and derived expansion velocity (420 \kms)
immediately suggested an evolution  similarity with He2-104, the "Southern Crab".
The 1986 low resolution and 1988 high resolution spectra presented there and in the
following papers (such as in CS93) pointed out significant
optical-UV spectral changes between 1986 and 1987, on a time-scale $\leq$ 1 year.
Since then, the link between SSs and bipolar planetary nebulae has become stronger,
demonstrating the need of improved models to explain such complex sets of data.

\section{The 1962 spectrum}

We start by analyzing the spectrum presented by HC80 and taken on 14 June 1962. We focus mainly
on the BI Cru emission lines in 1962, because
there are too few forbidden lines in the 1983 spectrum reported by R88 to allow a reliable modelling of  the nebulae.

In 1962, the complex and unusual spectrum includes HeII 4686 and blended NIII 4634,41,42, while [OII] is not visible.
HeI 4043 and 4026 show absorptions and P Cyg profiles. These lines led HC80 to suggest that
one component of the system possesses an
expanding atmosphere with an improbably large velocity of 1940 \kms, and to conclude that the
forbidden lines may arise in a region
apart from the one in which the permitted lines occur.

The analysis of the ion velocities shows three distinct velocity regimes: one at -55 \kms
represented only by HI, a second at -115 \kms which includes HeI and FeII, and a third at about -180 \kms
which includes HeII and the forbidden lines.
The individual velocities give an approximate mean error of $\pm$ 20 \kms. Only the  [OIII] 4959 line
shows a velocity of $\sim$ 377 \kms. In Table \ref{tab:t1}, the observed intensities corrected for
reddening, and the FWHM are shown for the forbidden lines and   He permitted lines.

Since the several emitting components recognizable in a symbiotic spectrum can be alternately visible or
hidden from view depending on the different configuration with respect to the line of sight,
as well as on the activity phase of the system, in the following we start identifying the emitting nebulae through their emission signatures.
Then, we try to interpret the spectra on the basis of the orbital motion and of the activity status of BI Cru.

\subsection{The colliding wind scenario}

In the last years, several observational evidences have proved that compact hot stars loose mass through fast (~1000 \kms),
low-density wind. Regarding SSs,  Nussbaumer et al. (1995) were the first that, by analyzing observations of a fast wind from the compact object,  called for a symbiotic colliding-wind scenario.

The winds from the hot and the cool stars  collide
within and outside the system, hence creating a complex
network of wakes and shock fronts that result in a complicated
structure of gas and dust nebulae (Nussbaumer 2000).
In the light of this scenario, one can consider that
two main shock-fronts develop from the head-on collision of the winds between
the stars. The \textit{binary} colliding wind configuration discussed in some theoretical works,
e.g. Girard \& Willson (1987), Kenny \& Taylor (2005), describes one strong shock-front facing the WD and the other, very weak, facing the cool giant.
Similarly, in the extended circumbinary region, two shock fronts develop
from the head-on-back collision of the winds: one expanding outwards
and the second, virtually negligible, facing the system center. This latter shock
network corresponds to
the \textit{concentric} colliding wind configuration of e.g.
Kwok (1988a), which is  a theoretical basis to the more realistic
picture of shock fronts disrupted by instabilities.

Generally, we can figure out that for all shock waves there is an \textit{upstream} region, where
the gas enters into the shock front, and a \textit{downstream} region, on the opposite side.
The shock front discontinuity is very thin, the thinner the faster the shock velocity. \\
The gas entering the strong shock front is thermalized and a high temperature region appears
immediately behind the
discontinuity, in the downstream region. On the opposite side of the shock front, upstream,
the temperature of the
gas is slightly increased by precursor radiation from the hot slabs of the gas downstream
and by radiation from the WD, without reaching temperatures as high as those in
the immediate post-shock region, though. The gas recombines following a high cooling rate due to the
high pre-shock densities and to compression downstream.

The key parameter is obviously the shock velocity, \Vs. For high-velocity shocks, the very
high temperature reached in the post-shock region leads to the X-ray emission observed
in several SSs. Moreover, broad  strong lines can also
be observed along the whole electromagnetic spectrum, particularly the coronal lines in the
infrared  (e.g. Angeloni et al. 2007a): therefore,
from the spectral point of view, different line profiles trace different velocity regimes,
allowing to highlight the different physical conditions within a symbiotic system.

Also BI Cru shows the signatures of the colliding-winds, because the range of velocities
observed in literature (Sect. 2.2) fits with both the head-on shock between the stars
("high-velocity" regime) and the expanding shock front outward the system ("low-velocity" regime),
and cannot be reconciled with a single velocity model. For instance, the [OIII] 4959 line observed
in the central system corresponds to $\sim$ 400 \kms, while [NII] lines observed in the lobes display
widths of $\sim$ 200 \kms. Lower velocities ($\sim$ 50 \kms) of optical-UV lines appear also in the
1962 spectra reported in HC80.

In the frame of colliding winds    already exploited to interpret many others symbiotic spectra
(Contini et al. 2009b and references therein), we schematically refer to the nebulae
downstream of the main shock fronts: 1) the head-on shock between the stars facing the hot star
(named hereafter \textit{reverse shock}) and 2) the head-on-back shock propagating outward the system
(hereafter \textit{expanding shock}). Moreover, we also consider 
3) the shock front accompanying the expansion of the lobes (Sect. 5).
The nebula downstream of the reverse shock between the stars is thus characterized by photoionizing
radiation and shocks acting on the same  edges of the shocked nebula,
while the models describing the expanding shock propagation outside the system  are characterized by
photoionization and shocks acting on opposite edges. For both the inverse and expanding shocks,
the shock velocity is suggested by the FWHM of the line profiles, while
the other  physical conditions in the emitting region downstream  are constrained by the observed line ratios.

The models must then account consistently for shocks and photoionization.
This is one reason why the SUMA code is particularly suited for this kind of spectral analysis applied to SSs.
The input parameters of the code are those relative to the shock: the pre-shock density \n0,
the shock velocity \Vs, the pre-shock magnetic field \B0; and those relative to photoionization:
the hot star ionizing radiation flux, its color temperature \Ts, and the ionization parameter $U$.
The chemical abundances of He, C, N, O, Ne, Mg, Si, S, Ar and Fe, relative to H,
the dust-to-gas ratio $d/g$ and the geometrical thickness of the nebulae $D$, are also accounted for.
Notice that $D$ is a lower limit in radiation-bound models.
Dust reprocessed radiation and bremsstrahlung  are consistently calculated throughout the nebulae,
as well as the dust  grain sputtering.

Specifically, the line and continuum fluxes downstream are calculated integrating throughout many plane-parallel slabs (up to 300)
with different geometrical widths  derived automatically   from the
temperature gradient.
By the way, the plane-parallel geometry is necessary
in the interbinary region where the collision of the winds is head-on.
It is also valid in the circumbinary region where the radius of the expanding shock
created by the head-on back interaction of the outflowing winds from the stars is large enough.

Furthermore, since the matter is highly inhomogeneous at the shock fronts because of instabilities at the nebula interface
(e.g. the Rayleigh-Taylor (R-T), Kelvin-Helmholtz (K-H), Meshkov-Richtmyer (M-R) instabilities), different physical conditions
should be accounted for, particularly regarding  the density.

\begin{table*}
\begin{center}
\caption{Optical line ratios to \Hb\ and model parameters. \label{tab:t1}}
\begin{tabular}{llllllll}\\ \hline  \hline
  line & Vel.&obs$^a$ & m1   & m2   &m3 &m4   & m$_{av}$\\ \hline
\ [NeIII]3869+3896  &-152& 0.17 & 0.26&0.004& 0.154&0.06 & 0.113    \\
\ [SII] 4068+4077 &-174& 0.03 & 0.001&0.0&0.3 &0.008& 0.02  \\
\ [OIII] 4363 &-189& 0.09 &0.11&0.001 & 0.15& 0.09& 0.06   \\
\ HeI  4471  & -79 & 0.17 &0.048&0.11& 0.89&0.05& 0.13   \\
\ HeII 4686  &-227& 0.13 &0.0016&0.2&5.e-4&0.005 & 0.11   \\
\ [OIII] 5007+4959&-480&  0.33 &0.6&0.043&1.6 &0.44& 0.35  \\
\ \Hb $^b$  & - &-    &1.37e4&6.2e-3&0.086& 1.0 &- \\
\hline
\  \Vs (\kms)& -&- & 190 &150& 70 & 400 &- \\
\ \n0 (\cm3) & - &-& 2e5 &1.6e3&3e4 & 1.e5&-  \\
\ \B0 (10$^{-3}$ gauss)& -&- & 1&1&0.1 & 1&-  \\
\ U          & - &-&15    & 25   &- & 1 &- \\
\ type$^c$       & -&- & RDo   &RDo& SD & RD&-  \\
\ $D$ (10$^{15}$ cm)   & -&- &1&1  & 5 &0.1&-  \\
\ log w       &- &-  & -7.0  &-0.045& -2.2   & -5&- \\
\ N$_{e}$  (\cm3)  &- & 1e6-1e7$^d$&-&-&-&-&-\\
\ \Te  (K)     &- &$\sim$5.5e4$^e$&-&-&-&-&-\\
\hline
\end{tabular}
\end{center}

\flushleft
$^a$ reddening corrected (from HC80);

$^b$ in [\erg];

$^c$ RDo: radiation dominated model with radiation flux and shock acting on opposite edges
of the nebula; RD: radiation dominated model with radiation flux and shock acting on the same edge;
SD: shock dominated model (U=0). See text for details.

$^d$ from HC80

$^e$ evaluated from the observed [OIII]5007+4959/[OIII]4363 (col. 3)
\end{table*}

\subsection{Modelling the line spectrum}

The modelling of a line spectrum is based on some basic points. The most
significant are the following:

1) recombination lines (e.g. \Hb, HeI, HeII) depend strongly
on the temperature of the star and on the ionization parameter.

2) line ratios of a single element from the same ionization level
but corresponding to a different quantum configuration
depend on the physical conditions
of the emitting gas (density, temperature, etc).

3) line ratios of  single elements from different ionization stages
depend on the ionization rates: radiative and/or collisional.
The radiative ones are strong at temperatures $\leq$ 10$^5$ K,
while the collisional ones are strong at high temperatures.
Therefore collisional ionization rates are important
when shocks are at work.

4) ratios of lines from different elements are strongly linked
to the relative abundances.

For a spectrum with a rich  number of lines, these rules
act together  and constrain the models.
Generally, in SSs the spectra from different nebulae must be
accounted for at the same time. The results will depend
on the relative weights adopted to sum up  single nebula spectra.

The spectrum from Bi Cru in 1962 shows the HeI and HeII line ratios to \Hb\
which are examples of condition 1) and the [OIII] 5007+4959/[OIII]4363
ratio which refer to condition 2); however, lines from different levels
refer to different elements leading to an uncertain modelling.
Moreover, each line is characterized by a different FWHM,
indicating that a pluri-nebula model must be adopted.

We have tried to complete the insufficient  informations derived from
the line ratios from the modelling of other SSs.
For instance, the model with  \Vs=400 \kms is accompanied by a
high preshock density ($\sim$  10$^5$ \cm3) because located between the stars.
R88 indicated a WD  temperature of at least $\sim$ 26500 K , Bohigas et al. (1989)
proposed a preshock magnetic field
of $\sim$ 10$^{-3}$ gauss in SSs similar to that of isolated giants. This value
 was confirmed
e.g. in the CH Cyg system by Crocker et al. (2001) and Contini et al. (2009a).
The nebula network throughout  SSs is further  complicated because each of the nebulae
is characterized by   relative abundances  suiting either those of the WD atmosphere or
those of the red giant (Contini 1997).
We discuss the abundances in the following: in our first trial we have used solar abundances
(Allen 1973).

\subsection{The selected models}

We have run a grid of models  covering reasonable ranges of all the input parameters in order to find the best and most consistent fit
of calculated to observed line ratios.

Four models are selected amongst the best fitting ones and are described in the bottom of Table \ref{tab:t1}.
For each of them, \Ts = 26500 K  and  a dust to gas ratio $d/g$= 4 10$^{-4}$ by mass are adopted.

Models labeled with RDo indicate a radiation-dominated case where radiation flux and shock act on opposite edges
of the nebula (\textit{expanding shock}).
Models labeled with RD deals with a radiation dominated case for which radiation flux and shock
act on the same edge (\textit{reverse shock});  the models labeled with SD are shock dominated (ionization parameter U=0).

The [OIII] 4959+5007 FWHM of 400 \kms  is accounted for by  model m4. 
The other input parameters
are purposely chosen in order to give a negligible contribution to the other lines which show narrow profiles.
In the frame of the colliding wind picture, model m4 represents the reverse shock front between the stars
which  explain the ISO IR spectra of D-type SSs, so we adopt \n0=10$^5$ \cm3 (Angeloni et al. 2007a).
Model m4 is radiation dominated and represents the shocked nebula downstream of the shock front facing the hot star.

The [OII]FWHM (200 \kms) indicate that a model with \Vs=200 \kms should be adopted (model m1). To further
constrain this model, we notice that such velocities are characteristic of the expanding shock. We adopt therefore
preshock densities of $\sim$ 10$^5$ \cm3  as, e.g., for  CH Cyg (Contini et al 2009b). 

Model m3 shows a very high HeI 4471/\Hb\ ratio and a low \Vs, in agreement with
the FWHM reported by HC80.
Lower velocities ($\sim$ 50-70 \kms)
are suitable to the expanding shock front at a large distance from the stars, therefore, model m3 is
characterized by U=0. 
Model m3 represents the shock dominated case, i.e.
  the ionization conditions throughout the nebula  are
dictated only by the shock, the flux being
absorbed   by some intervening matter (a dust shell?).

The geometrical thickness D are constrained by the dimensions of the system,
while the other parameters (e.g. U) are chosen phenomenologically for all the models.
Models m1, m3, and m4, which are dictated by the observations, either refer to a rather strong shock
or are shock dominated, leading to HeI/HeII $>$ 10. The observations show HeI/HeII=1.3, suggesting that
another   nebula dominated by a strong photoionizing flux,  should contribute to the averaged spectrum. 
Model m2 with U=25 produces a very high HeII/\Hb line ratio.

The models are  summed up adopting relative weights {\it w} which  lead to the fit of all the observed line
ratios at least within a factor of 2 (model m$_{av}$, Table \ref{tab:t1}) 
with a larger precision for the strongest lines. 

Complex models such as the pluri-nebula ones  used for BI Cru, require the combination of two calculation processes.
The first is achieved by the SUMA code which calculates the spectrum  emitted from a single nebula, the second is 
an {\it ad hoc} program which provides the weighted sum  of the single-nebula spectra. 
The results of the two processes are cross-checked
until a fine tune between the observed and calculated  line ratios is obtained.
The whole  procedure  requires  a 
large grid of models which are  constrained by the  data, by the range of the
 physical parameters in symbiotic systems, and by consistency  of line and continuum modelling. 

The average line ratios are calculated by:

\noindent
 (I$_{\lambda}$/I$_{H\beta}$)$_{av}$ =  $\Sigma_i$ ((I$_{\lambda}/I_{H\beta}$)$_i$ (I$_{H\beta}$)$_i$ w$_i$) / $\Sigma_i$(I$_{H\beta}$)$_i$ w$_i$

\noindent
where i= 1 to 4 refers to the models. 
The calculated line ratios  (I$_{\lambda}/I_{H\beta}$)$_i$ and  calculated absolute
\Hb fluxes (I$_{H\beta}$)$_i$ are given in Table 1.

The absolute fluxes of the lines  are very different  for the different models, depending particularly
on n$_0^2$. 
Mathematically, the weights w$_i$  roughly compensate for the line fluxes.  Physically, they
represent the $\eta$ factors which are introduced in Sect. 6 in order to compare fluxes calculated 
at the nebula with fluxes observed at Earth.

The results obtained by modelling the line spectrum lead to a better understanding   of BI Cru  system.
For instance, the  contribution of model m2, which is characterised by \Vs=150 \kms and \n0$\sim$ 10$^3$ \cm3,
  leads to a better fit of the HeI/HeII line ratio. 
This shock velocity is also suitable to the expanding shock, even if
models m1 and m2 show different pre-shock densities by a factor of $\sim$ 100, and a  different ionization parameter.
This  confirms that the expanding shock is very disrupted and propagates in the a
 non homogeneous medium.

Shock velocities through clouds of different densities are inversely proportional to the density square root ratio:
this is evident when comparing \Vs\ and \n0 of models m2 and m3.
In turn, model m3 shows both lower \Vs\ and \n0 than model m1, thereby indicating that the
corresponding nebula has reached
a larger distance from the central system than that corresponding to the model m1.

\begin{figure}
\begin{center}
\includegraphics[width=0.45\textwidth]{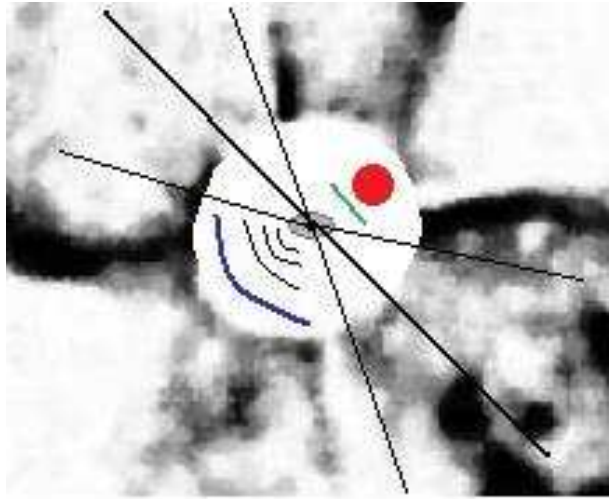} 
\caption{The BI Cru symbiotic  system as derived from
our modelling is sketched over the image observed by SC92.
The Mira and the WD encircled by the accretion disc
are recognizable, as well as the reverse shock front between the stars
(model m4, in green), the expanding shock fronts (m1, m2,and m3, in black),
and the shock front accompanying the blast wave from the WD outburst (blue).
\label{fig:sk}}
\end{center}
\end{figure}

A sketch showing the location of the  shock fronts within BI Cru symbiotic
system is shown in Fig. \ref{fig:sk}

Notice that   calculated [NeIII]/\Hb\ and [SII]/\Hb\
can be ameliorated  adopting relative abundances of Ne and S  slightly higher than
solar by a factor of$\sim$ 1.5.  We can deduce directly from the fitting results
the abundances of Ne and S because they are not strong coolant.
Ne/H and S/H higher than solar might be  characteristic
of WD  atmospheres,  although in BI Cru they are  highly  diluted
by merging with the ISM. 

The electron density \Ne\ and the electron temperature \Te\  measured from the
observations appear in the bottom of Table \ref{tab:t1} for comparison.

\begin{figure*}
\begin{center}
\includegraphics[width=0.4\textwidth]{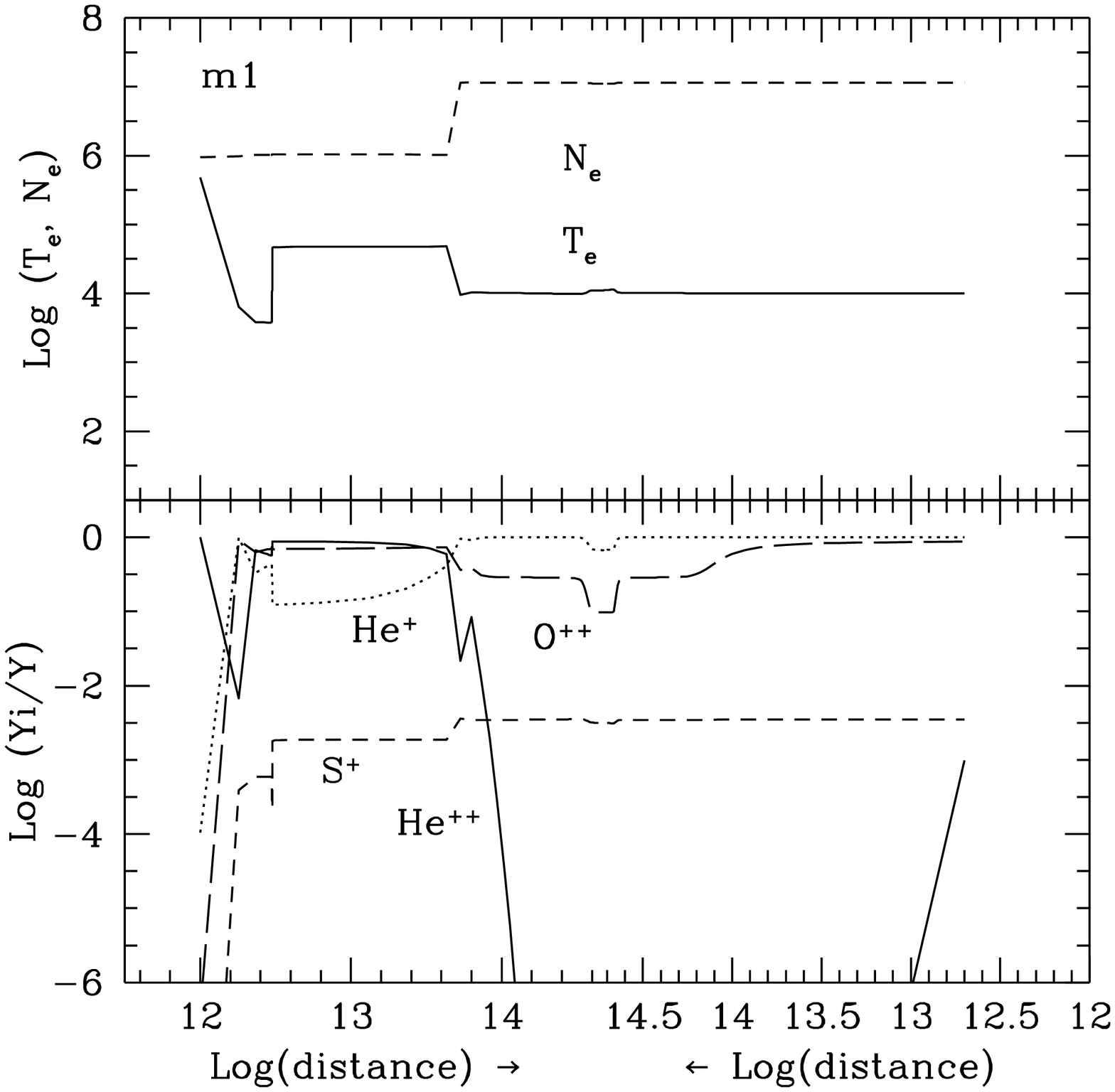}
\includegraphics[width=0.4\textwidth]{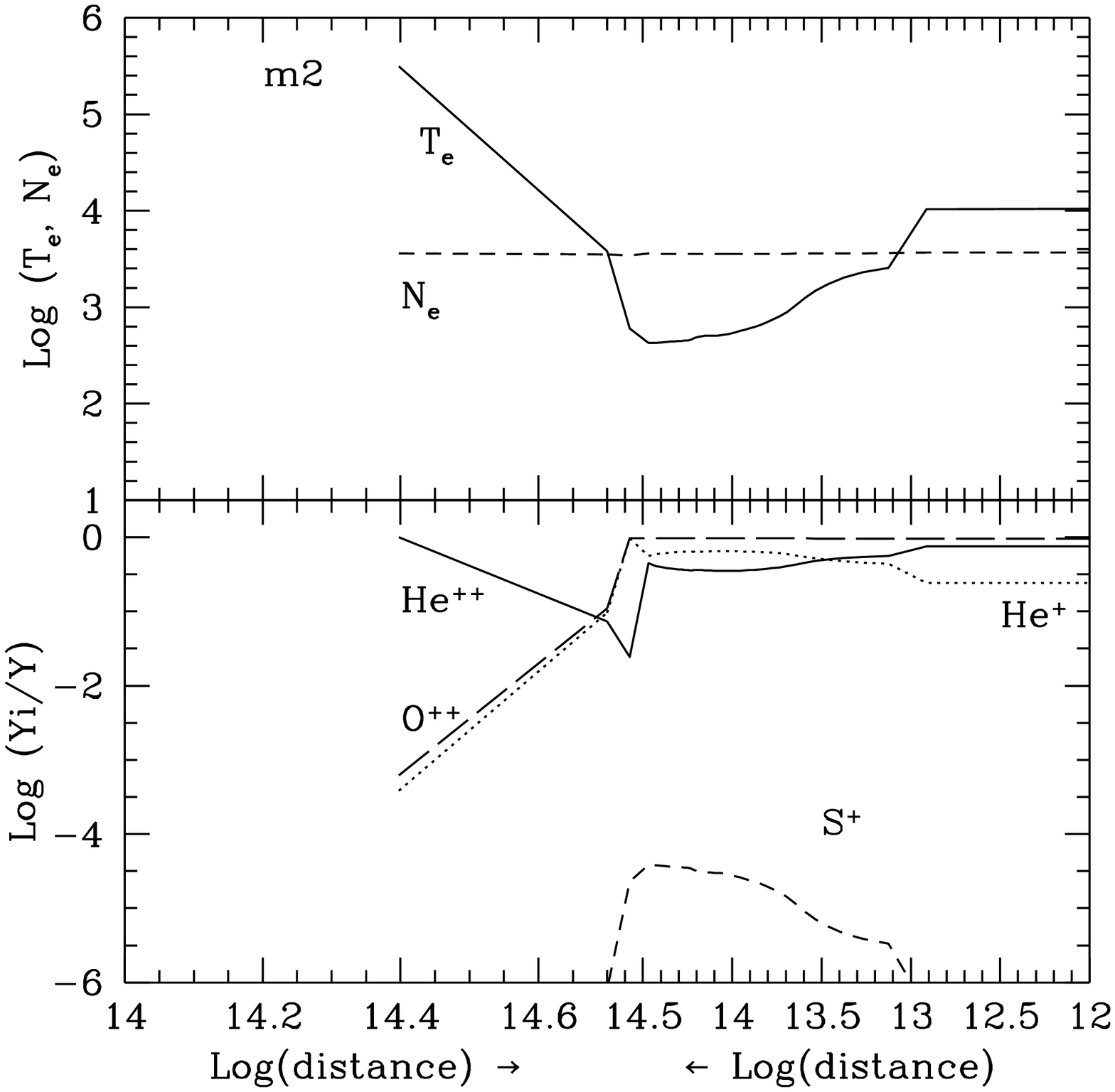}
\includegraphics[width=0.4\textwidth]{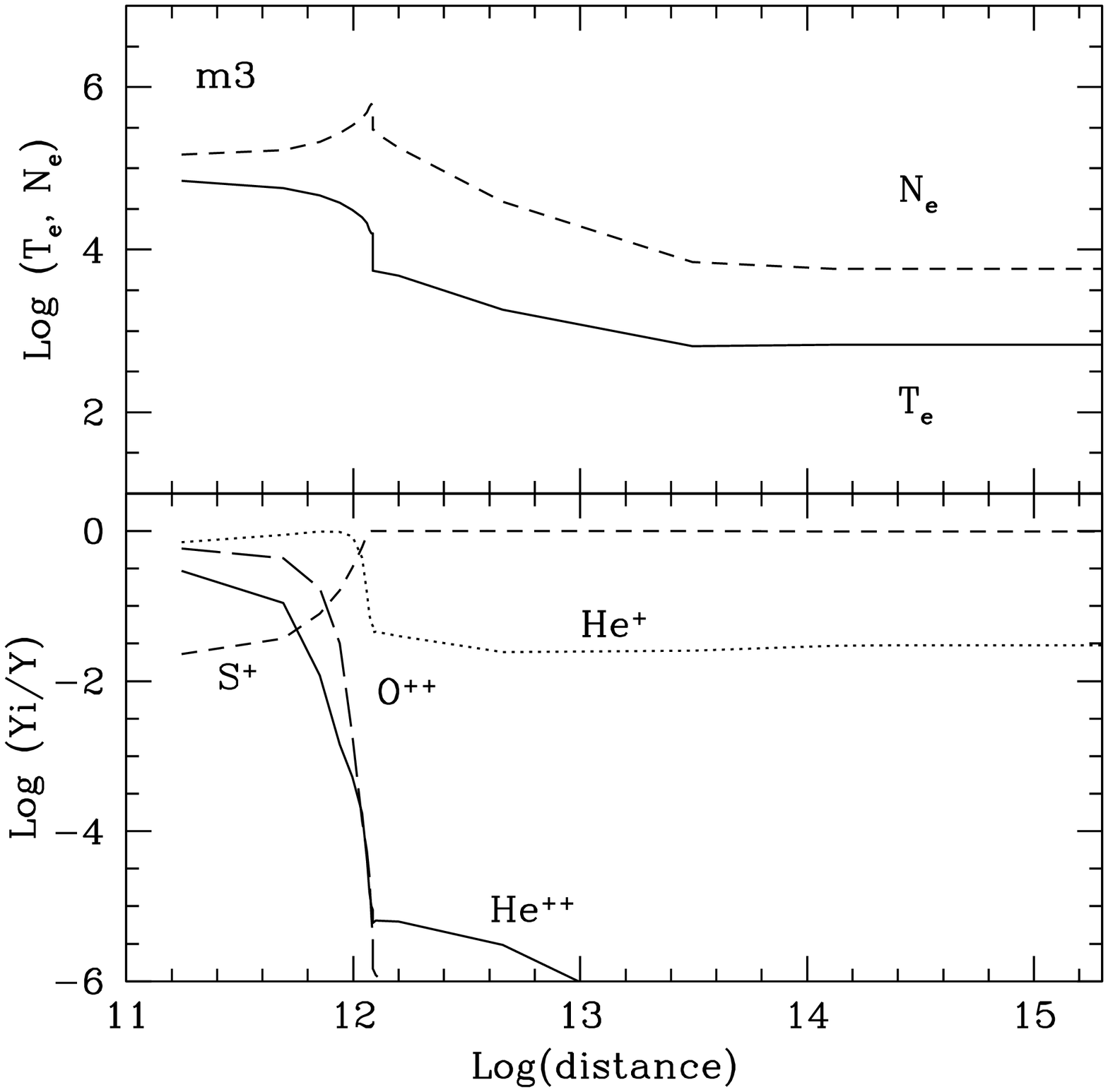}
\includegraphics[width=0.4\textwidth]{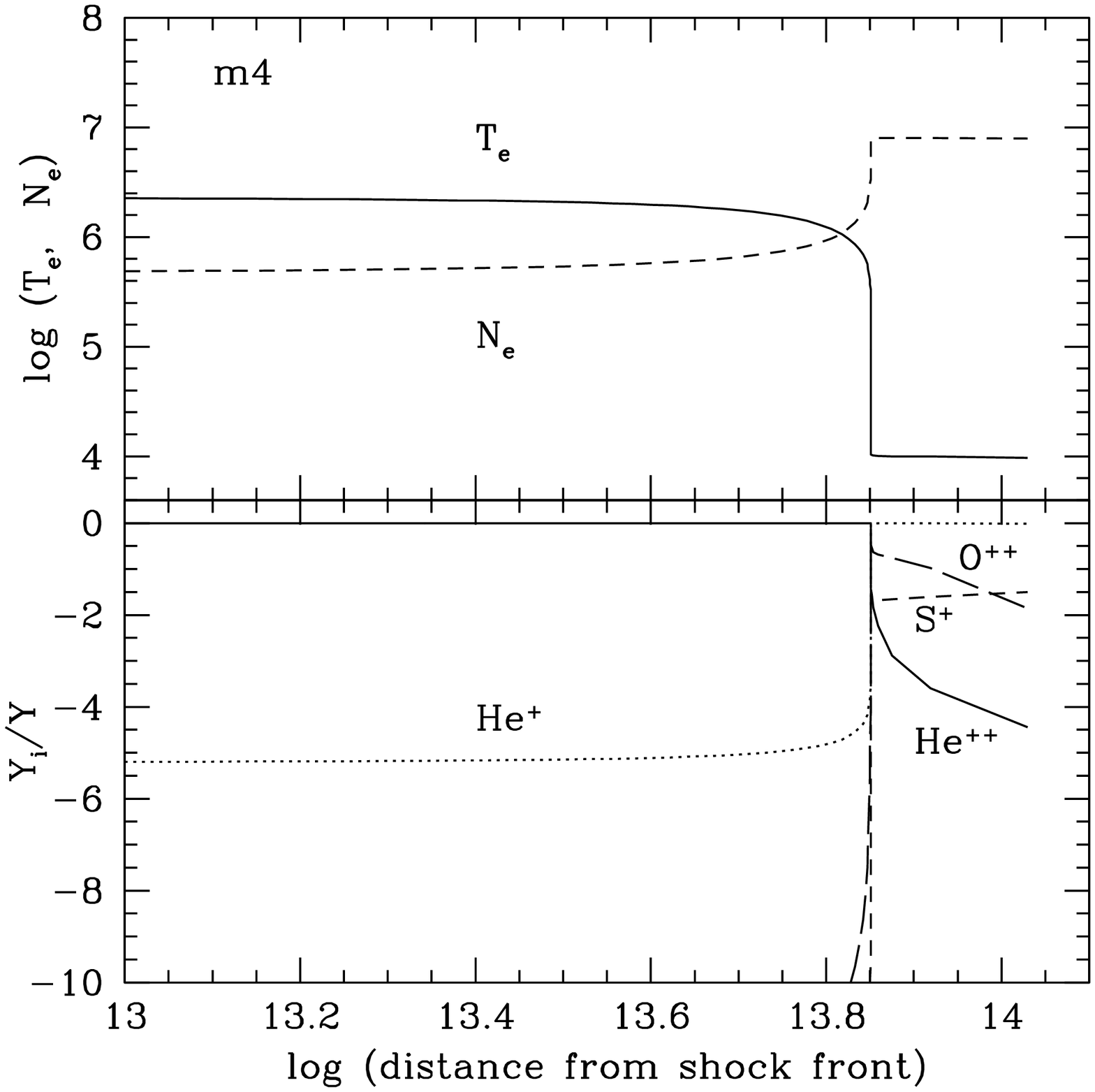}
\caption{The profiles of the electron temperature T$_e$, the electron density N$_e$ (top panels) and of 
the fractional abundances of the most significant ions (bottom panels) in the different models described in Table \ref{tab:t1}. 
For all diagrams the shock front is on the left and the distances  are in [cm]. 
Models m1 and m2 (top diagrams) show the case of shock and photoionization acting on opposite edges of the nebula: 
therefore the X-axis scale is symmetric with respect to the centre, in order to have a good resolution of the conditions  
in both the edges of the nebula. The scales however are different as requested  by the calculation accuracy. 
In this way,  shock dominated region (on the left) and radiation dominated region (on the right) are  clearly recognized. 
The two regions are not independent because bridged by the secondary radiation. 
The two diagrams on the bottom of the figure represent a shock dominated model (left - m3) and the case that shock and 
photoionization act on the same edge of the nebula (right - m4).\label{fig:prof}}
\end{center}
\end{figure*}

Using SUMA, the spectra emitted from  each nebula  result  from
integration throughout  different gas regions downstream
characterized by different physical conditions which  derive from the cooling rate,
from  radiation transfer
of the primary and secondary (diffuse) radiation flux, and from compression which characterizes
models accounting for  the shocks.
We present in Fig. \ref{fig:prof}  the profiles of the electron density N$_e$, electron temperature T$_e$,
and of the fractional abundance of the most significant ions which lead
to the different lines ratios.
The lines in Table \ref{tab:t1} correspond to low and intermediate
ionization levels, meaning that most of the lines are emitted from gas at $\leq$ 5 10$^4$ K.

\subsection{The FeII lines}

Finally, we refer to the several permitted FeII lines, often recorded in BI Cru
spectra (Fig. \ref{fig:rossi}) but not included in Table \ref{tab:t1}, which  deserve a special discussion.
Emission lines of FeII are seen in the ultraviolet spectra of many  SSs (e.g. RR Tel, AG Peg).
In some objects, optical multiplets of FeII are also seen in emission, indicating that the ultraviolet
resonance lines are optically thick (Penston 1987).

Transitions between even 5-eV levels and even 3-eV levels correspond to the permitted optical
multiplets 27, 28, and 29, around 4000 \AA, and to the well-known feature at $\sim$ 4570 \AA\ (Collin \& Joly 2000),
which are all observed in BI Cru, while forbidden [FeII] lines are rarely observed. Therefore,
 an overabundance of iron cannot help to explain the FeII emission
by a photoionization model,  but the FeII region should be heated by an additional mechanism;
that is, the FeII spectrum is not produced directly by photoionization but more probably by shocks.
Indeed, it is generally believed that collisional excitation is responsible for the bulk of the FeII emission, 
and it has been shown how these lines may represent, especially in SSs, one of the most direct traces of fast 
outflows of WD winds (Eriksson et al. 2007).
Inelastic collisions with electrons excite the odd parity levels near 5 eV which then decay into the optical and UV lines.
Self-fluorescence and fluorescent excitation by \Ly are also important sources of excitation (Sigut \& Pradhan 2003).

Verner et al. (2000) have shown that at low densities (n$_e$ $ \leq$ 10$^2$ - 10$^4$ \cm3) the permitted
optical Fe II lines are relatively weak, the reason being that the 63 lowest levels, the most populated
at these densities, are all of the same (even) parity and are able to radiate only forbidden lines.
The situation dramatically changes near 10$^6$ \cm3 because, then, levels of odd parity are populated
by collisions, enough to produce the permitted lines. Therefore, if both the permitted Fe II  and forbidden
lines were  produced in the same region, the density should be larger than 10$^6$ \cm3 and lower
than 10$^8$ \cm3  because forbidden lines would be collisionally deexcited
(Veron-Cetty et al. 2004).

In the  case of BI Cru, it is interesting to note that models m1 and m4 correspond to 2 10$^5$ \cm3
and 10$^5$ \cm3, respectively, which lead to n$_e$ $>$ 10$^6$ \cm3 downstream. In this density range,
the FeII lines can then be produced without invoking a different emission region for permitted and forbidden
lines, conversely to what stated by HC80.

\section{The broad \Ha\ line}

\begin{figure}
\begin{center}
\includegraphics[width=0.45\textwidth,angle=-0.4]{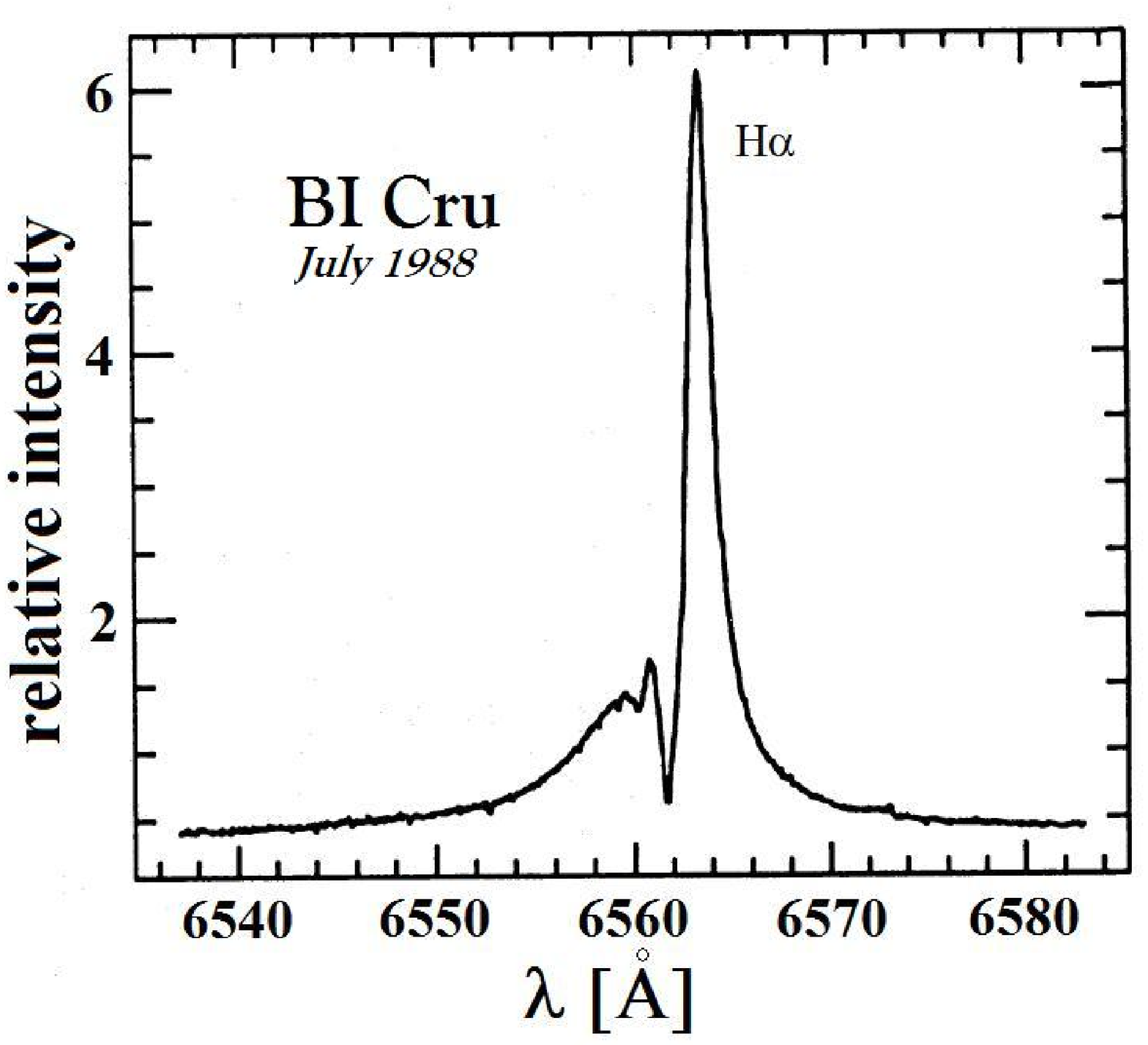}
\includegraphics[width=0.45\textwidth]{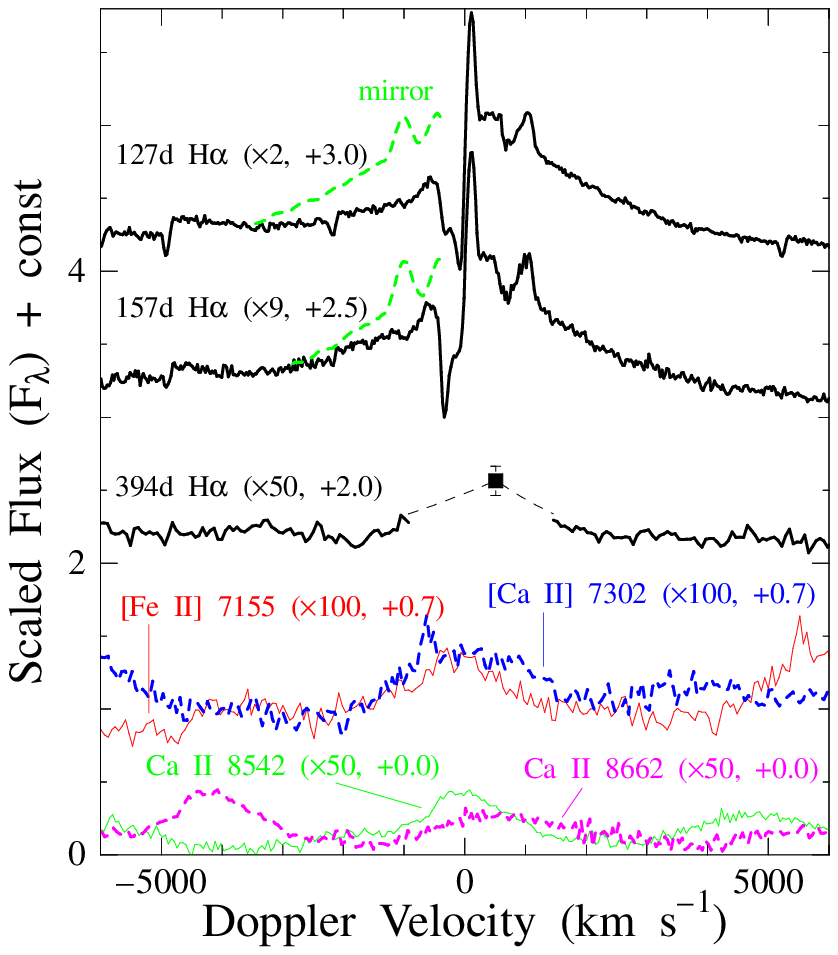}
\caption{Top: High-res spectrum of \Ha as recorded in July 1988 (adapted from SC92).
Bottom: The \Ha spectrum observed from the supernova 2006gy (taken from Kawabata et al. 2009) \label{fig:Ha}}
\end{center}
\end{figure}

Whitelock et al. (1983) reported on a strong \Ha with blue displaced P Cyg absorption at -228 \kms
and extensive emission wings with FWHM of $\sim$ 1500-2000 \kms, observed in the 1974 spectra.
On Feb 1984, R88 observed the \Ha region, noting a double emission with a strong and broad violet-shifted
absorption extending to about -300 \kms. The \Ha absorbed portion was thought to arise from gas in
front of an HII region with a large velocity gradient, hence suggesting a possible location in the
cool giant wind accelerated by the intense radiation of the hot component.
SC92 recorded, on a high-resolution spectrum taken in 1988, an \Ha line with a FWZI of more than
3000 \kms (Fig. \ref{fig:Ha}): such velocities are generally explained by scattering in an accretion
disk (Robinson et al. 1989).

Linear spectropolarimetry of BI Cru was presented by Harries (1996).
Interestingly, he found that the broad blue \Ha wing is unpolarized, while the red one is
strongly polarized. As already suggested by R88, Harries (1996) proposed that the \Ha\ emission is produced
in two separate velocity regimes: the central narrow peak being
formed in the slow moving part of the cool component wind, while the broad component in the part
of the wind approaching the hot source.
The spectropolarimetric observations supported this hypothesis, although Harries (1996) believed the source of the high
velocity material had to be identified in a bipolar flow.

The broad blue wing of \Ha could neither  be emitted from the accretion disk
because the blue wing and the red wing show different polarization.
 Actually, lines emitted from an accretion disk are generally double peaked with asymmetry in peak heights. 
The modelling by Robinson et al. (1994) leads to two  peaks with similar intensity, while
 in SSs that are believed to contain an accretion disk, eclipses are invoked to explain unequal
emission peaks. 
For instance, an exceptionally  broad nuclear \Ha was observed in M2-9 by Balick (1989) and explained as emitted from
the accretion disk because double-peaked. However,  the ratio of the red and blue component intensities  and
 a line width of 11,000 \kms    deserve a different interpretation.

Our analysis of  the exceptionally broad \Ha in BI Cru focuses on the evidence that the blue wing is unpolarized, meaning that
this part of the line is formed out of the scattering region, and
  excluding therefore    broadening by scattering  by a high opacity (Mikalojewska et al. 1988).

Moreover,  in Fig. \ref{fig:rossi} the \Ha line appears abnormally broad, whereas the other
strong permitted lines (e.g. He I 5876 and 6676) which should be emitted from the same emitting gas region,
are both much narrower.

According to our recent interpretation concerning   the appearance  of a broad \Ly in the CH Cyg spectra at the
end of the active phase 1977-1986 (Contini et al. 2009b), we would like to explain  the broad \Ha line in BI Cru
by means of the theory of charge transfer reactions between ambient hydrogen
atoms and post-shock protons at a strong shock front (Heng \& Sunyaev 2008).
Particularly,  recall that in the circumstellar side of the WD opposite to the red giant,
the effect of symbiosis is less enhanced. Here, we can apply to the WD outburst the theory developed
by Chevalier (1982) for Type II supernovae.

We rely on the hypothesis of CS93 that multiple bursts could occur in BI Cru, similarly to He2-104.
The WD temperature $\sim$  26500 K found by modelling the line spectra is an indication that the
last burst is completely run out.
 Whitelock et al. (1983) reported on  FeII narrow lines
and  Allen's (1984) spectrum is rich in NIII and HeII lines which
 could be emitted from  the expanding region.
 According to the shock front network created by  collision of the wind, 
the expanding shock front is located in the circumbinary
side of the system opposite to the red giant star.
Therefore we  can  apply Chevalier's theory.

Actually, Kawabata et al. (2009) published the optical spectra of Type II SN 2006gy at late state detection.
A strong similarity with BI Cru  \Ha line
profile (Fig. \ref{fig:Ha}) can be noticed.
 
The interaction of the freely expanding matter with the surrounding medium gives rise to a high-energy
density region bounded by shock waves. Two shock fronts develop, one proceeding inward in the high density
region, the other expanding outward in the circumstellar medium.
Following Chevalier, we  consider the interaction of the outburst with circumstellar matter
on the assumption that it is built up by a steady wind. If the ambient density is described
by $\rho$ $\propto$ r$^{-s}$ ($\rho$= 1.4 m$_H$ n, where n is the density  in number \cm3 and
m$_H$ the mass of the H atom) the steady wind corresponds to \textit{s}=2.

By 3D hydrodynamic simulation  in the case of RS Ophiuci, Walder et al. (2008) found
that the density decreases throughout the nova remnant as 1/r$^2$ \textit{in average}. 
Accordingly,  we  will use
the case of the uniform expanding gas  described by \textit{s}=2 and $\gamma$=5/3.

For  \textit{s}=2, the radius of the outer shock corresponding to the blast wave 
R$_{BW}$
 is given by the Primakoff solutions (Chevalier 1982, eq.5) :
\begin{equation}
 R_{BW} = (3 E/2 \pi A)^{1/3} t^{2/3}  
\end{equation}
where E is the total energy (twice the kinetic energy),  $\rho_0$ = A R$_{BW}^{-2}$,
 $\rho_0$= 1.4 m$_H$ \n0,   and \n0 the density of the gas
upstream.

This equation is valid for times longer than a specific time, called the time of change
\begin{equation}
 t_{s} = 0.677 M_{ej}^{3/2} /A E^{1/2}
\end{equation}

between that of the interaction of freely expanding matter  with the surrounding medium and the
following one, i.e. when the flow tends toward the self-similar solution for a point explosion
in a power-law density profile (Sedov 1959). M$_{ej}$ is the ejected mass.

The velocity of the blast wave is:
\begin{equation}
  V_{BW} = dR_{BW}/dt = 2/3 (3E/2 \pi A)^{1/3} t^{-1/3}
\end{equation}

On the basis that no burst has been recorded in the last 60 years of observations, we adopt t$\sim$60 years as a lower limit.
 Following the method indicated  by  Contini et al. (2009b) for CH Cyg,
a present blast wave velocity   V$_{BW}$= 1500-2000 \kms,  at least 60 yr after the outburst, would then translate to
V$_o$ $\sim$  5800-7700  \kms  one year after the burst.
 A velocity of $\sim$ 5800 \kms is about the maximum predicted by the models of Yaron et al. (2005)
for nova outbursts.  A period shorter than one year after Bi Cru burst would lead to higher velocities.
  We cross-check whether  one year from the burst
is compatible with the time of change t$_s$ (eq. 2). If so, we adopt an initial
velocity of 5800 \kms as the escape velocity.

 The results of   Yaron et al (2005) and Prialnik \& Kovetz (1995) models
of nova outbursts indicate 
that a velocity of 5800 \kms is possible for a   $\geq$ 1 \msol WD.

Considering  an  escape velocity of 5800 \kms  and  a WD mass M$_{WD}$=1. \msol,
we obtain  the radius of the WD, R$_{WD}$ $\sim$   6.7  10$^8$ cm.
 This in turn  corresponds to L* $\sim$  7 10$^{35}$ erg s$^{-1}$  for  \Ts=200,000 K,
still  below the Eddington luminosity.
Therefore,  a stellar wind  could not develop from BI Cru and confirms the hypothesis
of a nova-type wind.

The radius of the  blast wave in BI Cru
 60 years  after the outburst, is calculated  by  R$_{BW}$ = 3/2 V$_{BW}$  t, leading to 
R$_{BW}$=4 10$^{17}$ cm.
The ejected mass   is then calculated by   eq. 2.
Notice that A = 1.4 m$_H$ \n0  R$_{BW}^2$ is a constant. We assume that the preshock density \n0 at a radius of
4 10$^{17}$ cm from the binary system is of the order of that of the ISM (0.1 - 10 \cm3).
Then,  if t$_s$ = 1 year, M$_{ej}$/ \n0 = 4.8 10$^{-6}$ \msol cm$^3$, in agreement with 
 Prialnik \& Kovetz (1995).  

We can calculate by A  that a density of 10$^9$ \cm3 could be found
within  a radius of $\leq$ 1.2 10$^{13}$ cm from the WD.

\section{The bipolar lobes}

The central image  of BI Cru taken by SC92 (Fig. \ref{fig:CS92}) is in the [NII] light.
No [OIII] and  [OII] appear in the lobes. In the 1986/87 spectra they have identified:
HI,HeI,HeII,OI,OII,OIII, [OIII],NII,NIII,[SII], SiII, FeII,[FeIII],[FeVI], [ClIII],[ClIV].
The 1987 spectrum, much richer than the 1986 one, confirms the important variability of the
object.

\begin{figure}
\includegraphics[width=0.48\textwidth]{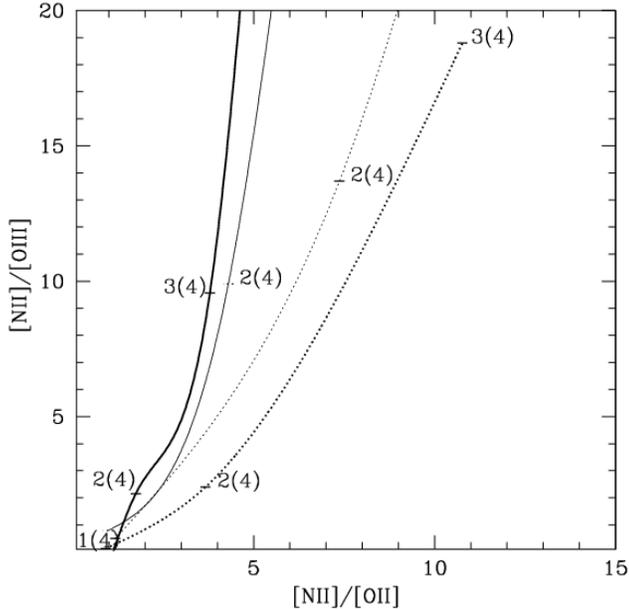}
\caption{[NII] 6584+6548 / [OIII]5007+4959 vs [NII] 6584+6548 / [OII] 3727+3729 for different models.
Thick lines: \Vs=250 \kms; thin  lines: \Vs=200 \kms; solid lines: \B0=10$^{-4}$ Gauss;
dotted lines: \B0=10$^{-3}$ Gauss. Labels refer to \n0 in 10$^4$\cm3. \label{fig:el}
}
\end{figure}

\begin{figure}
\includegraphics[width=0.45\textwidth]{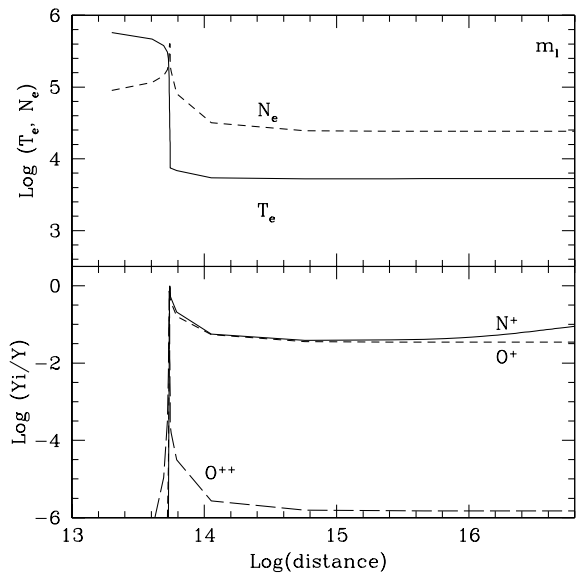}
\caption{The profiles of T$_e$, N$_e$ (top panels)
and of the fractional abundances of the most significant ions
(bottom panels) for model m$_l$, representing the conditions in the lobes. \label{fig:lobes}}
\end{figure}

The ejecta (jets) interact with the ISM with velocities  $\leq$ 200 \kms,
similarly to what happens in the He 2-104 'crab legs'. Two lobes are seen in the [NII] light  but not
in  the  [OII] and the  [OIII], in both  BI Cru  and He2-104.
Contini \& Formiggini (2001) have  modelled and discussed the formation of
the "crab legs" in He2-104: therefore we follow the same modelling approach.
First, we constrain the models by [NII]/[OII] and [NII]/[OIII] $\gg$ 1.
We have run a shock dominated (U=0) grid of models with \Vs\ between 200 and 250 \kms
as measured by SC92. Fig. \ref{fig:el} shows that the critical line ratios result easily for \Vs =200 \kms.
Higher velocities can be excluded.
The pre-shock densities are $\geq$ 10$^4$ \cm3, high enough to reduce the [OII] line
intensity by collisional deexcitation downstream,  as can be seen in Fig. \ref{fig:lobes}, where
the electron density \Ne,  electron temperature \Te, and the fractional 
abundances of the most significant ions in the lobes, are shown  
as a function of distance from the shock front (in cm).
 The magnetic field is \B0=10$^{-3}$ Gauss,
the same as that found in the system centre. The geometrical thickness of the filaments in the lobes is
$D$=10$^{17}$ cm, leading to radiation-bound models.

\section{The continuum SED}

\begin{figure*}
\begin{center}
\includegraphics[width=0.48\textwidth]{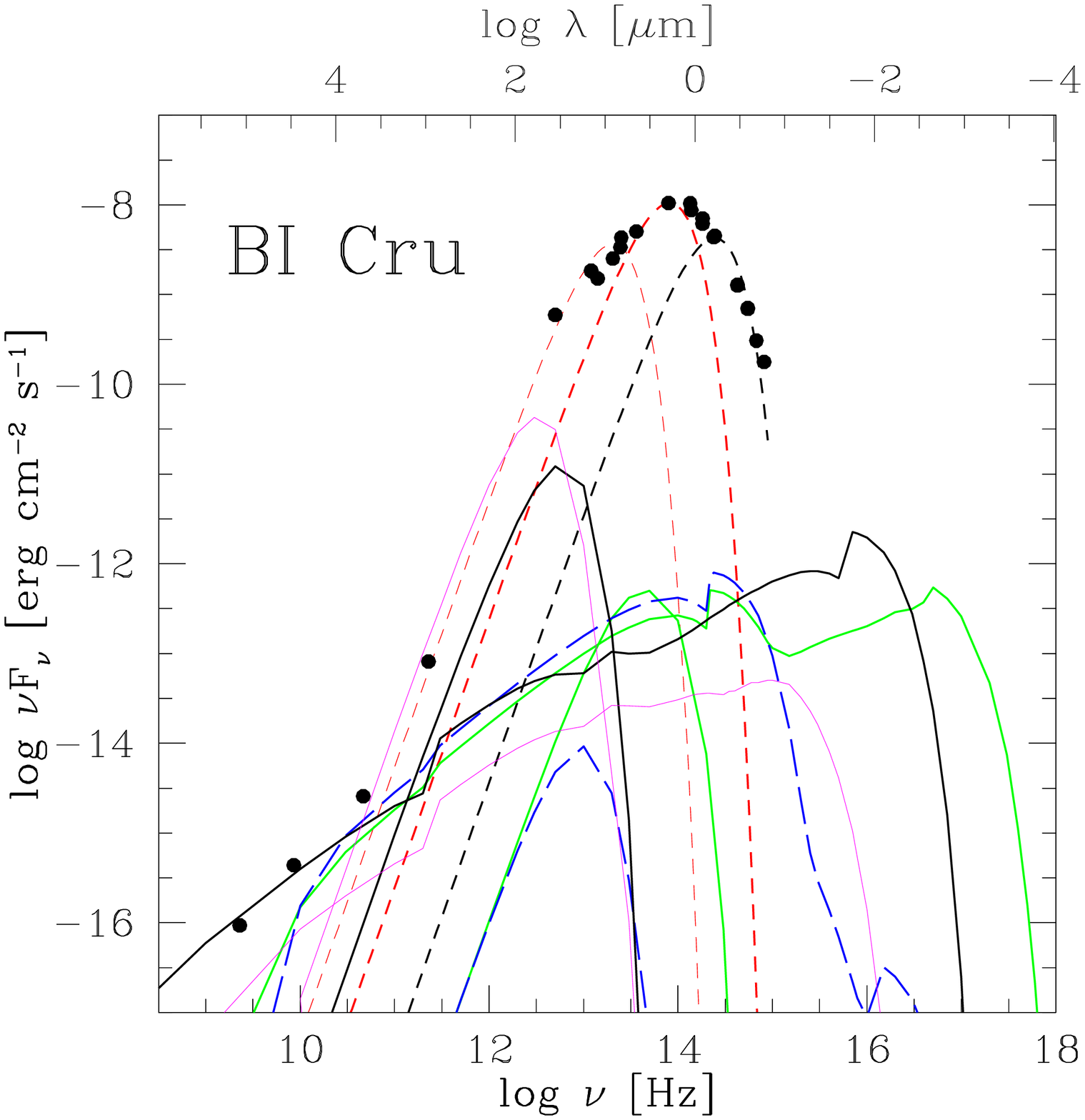}
\includegraphics[width=0.48\textwidth]{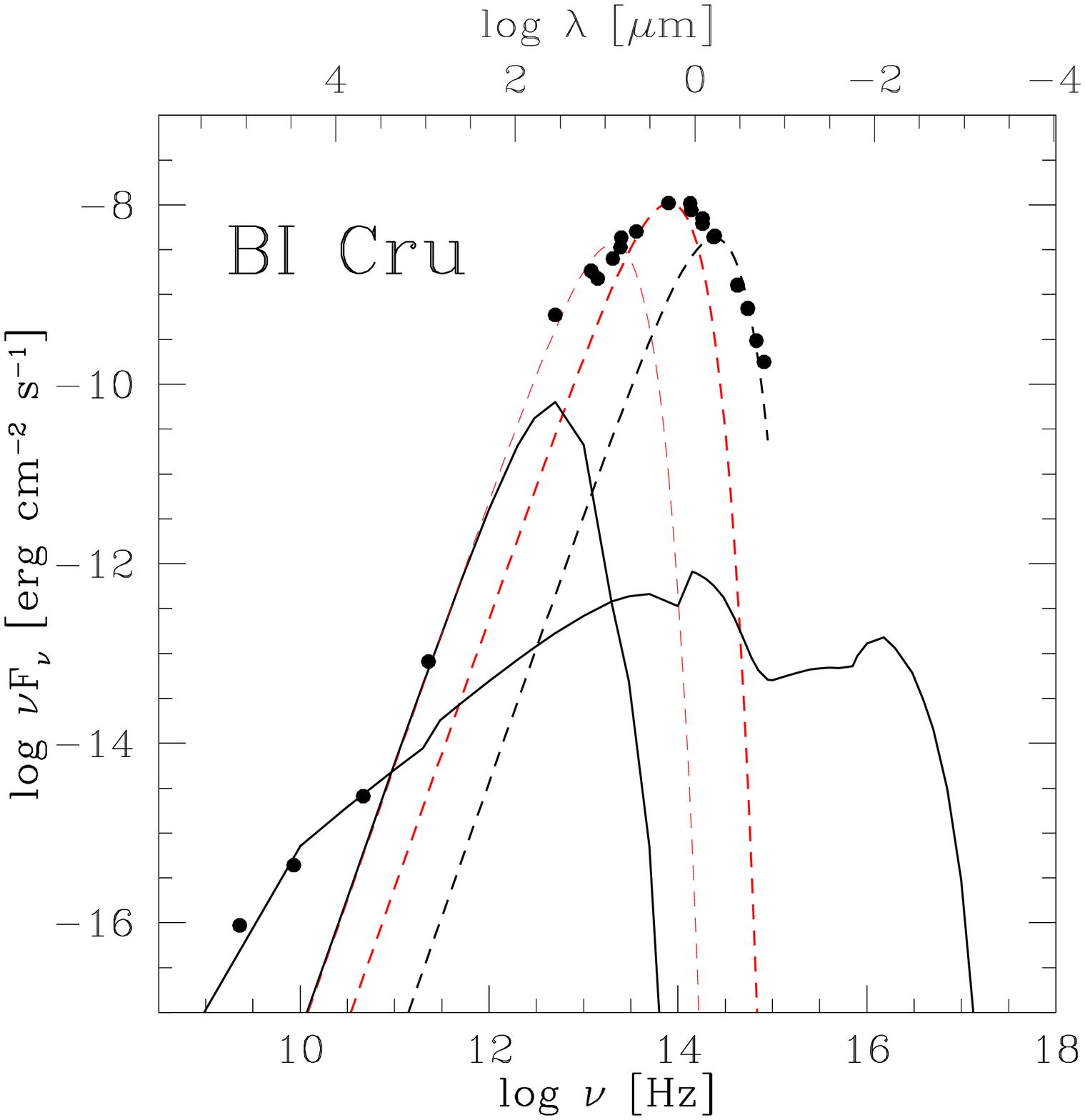}
\caption{The observed BI Cru continuum SED, from radio to UV.  The data (filled circles) are described in Sect. 2.
Thin short-dashed line (red): 250 K dust shell; short-dashed line (red): 850 K dust shell;
thick short-dashed line (black): Mira component.
Left diagram:  solid lines (green): m4; long-dashed lines (blue): m1; thick solid lines (black): m2;
thin solid lines (violet): m3. Both bremsstrahlung and re-radiation by dust are shown for each model.
Right diagram: solid lines (black): model m$_l$. }
\label{fig:sed}
\end{center}
\end{figure*}

The observed  SED of the continuum, calculated consistently with the line spectra
that were  extensively presented in the previous sections, is shown
in Fig. \ref{fig:sed}. There are no data for the continuum flux in the years corresponding
to the 1962 line spectra (Sect. 3).
We are thus forced to refer to later data, having already checked that the intrinsic flux variations
at different epochs are not as strong as to invalidate the results (Sect.2). At high frequencies
($\geq$ 10$^{15}$ Hz) there are no data, so we
 constrain the models on the  basis of the radio-IR-optical data presented in Sect. 2.

As we have shown in previous papers (e.g. Contini et al. 2009a), a few schematic models can describe
the variable spectra of SSs: therefore we  compare the combination of  models   calculated
by the  fit of  the line spectrum on the basis
of a data-set taken on a certain time,  with a mosaic of continuum  data
observed at very different epochs.
The modelling  is presented in Fig. \ref{fig:sed}, left diagram. The models  are scaled according to the
weights shown in Table \ref{tab:t1}, last row.

Model m4 (Fig. \ref{fig:sed}, left diagram, green lines) shows the characteristic self-absorption in the radio range
 already found in other SSs (e.g. H1-36, Angeloni et al. 2007b). It describes the physical conditions of the nebula
downstream of the reverse shock which is characterized by relatively high densities
downstream ($>$ 10$^6$ \cm3).  This model shows that X-rays could be expected in 1962.
Emission in the UV - soft X-ray  is suggested by model m2 (black lines) which explains also the data at longer
wavelengths. This model represents the nebula expanding with \Vs=150 \kms . The bremsstrahlung emitted
from the  shock dominated model m3 (violet lines)  is not  directly seen throughout the SED;
however, its reprocessed emission from dust may 
contribute to very far IR emission ($\nu$ $<$ 10$^{12}$ Hz).

We suggest that the presence of many shock fronts in BI Cru should be confirmed  by
synchrotron radiation in the radio range,  produced by the Fermi mechanism.

The line spectra are generally modelled referring to the line $ratios$, while the SED of the continuum is
modelled on the basis of $absolute$ fluxes.
Since the observations are taken  at the Earth, while the models are calculated at the nebula,
we define the factor $\eta$  =( \ff ~ R/d)$^2$, where r is the  distance of the nebula
from the SS center, d the distance to Earth ($\sim$ 2 kpc for BI Cru),
and \textit{\ff} the filling factor.
The $\eta$ factors, depending on  the distance of the nebulae from the system center,
further constrain the models.

Adopting a continuum SED  similar to that observed in later epochs,
we find that the radius of the different nebulae which contribute to the 1962 line spectrum
are  r$_{m1}$ = 6. 10$^{13}$ cm, r$_{m2}$ = 1.8 10$^{17}$ cm, and r$_{m3}$= 1.5 10$^{16}$ cm,
adopting \ff =1.

The distance of the reverse shock from the hot star is r= 2.8 10$^{13}$ cm, considering
that for model m4, U=1 can be
combined with  \Ts\  by means of $F_{\nu}$ (r$_{WD}^2$/r$^2$)=U n c, where $F_{\nu}$ is the flux in
number of photons cm$^{-2}$ s$^{-1}$ corresponding to  \Ts = 26500 K, and adopting r$_{WD}$=
5.4 10$^8$ cm (Sect. 3.3).

In Fig. \ref{fig:sed}, right diagram, we refer to the modelling of the lobe spectra (Sect. 3.3).
We select a  model (m$_l$) representing the filaments in the lobes, with \Vs=210 \kms as indicated by SC92,
\n0=2 10$^4$ \cm3, \B0=10$^{-3}$ Gauss, $D$=0.33 pc.
The bremsstrahlung calculated by model m$_l$  is then compared with the data in Fig. \ref{fig:sed} (right diagram,
blue lines), leading to  log $\eta$=-12.5.  SC92 indicate that the lobes had expanded to a radius  r$_l$$ \sim$1.3 pc.
Combining $\eta$ with r$_l$   we obtain \ff $\leq$ 0.001.\\

According to its nature of \textit{dusty} SS, also BI Cru confirms that in order to reproduce the NIR-MIR
continuum slopes, different ''dust'' temperatures should be combined (e.g. Anandarao et al. 1988,
Angeloni et al. 2009, in preparation). In BI Cru, two dust shells result from the continuum SED modelling,
an internal one at a temperature of 800 K with a radius of 1.5 10$^{14}$ cm, and the outer one at 250 K with a
radius of 1.6 10$^{15}$ cm, probably circumbinary.

\begin{table*}
\caption{Comparison of BI Cru, He2-104, R Aqr, and CH Cyg. \textit{Rev} and \textit{exp} stand for reverse and expanding shock, respectively. \label{tab:t2}}
\tiny{
\begin{tabular}{llllllllllllllllllll}\\ \hline  \hline
\  & \multicolumn{3}{c}{BI Cru$^a$}&&& \multicolumn{3}{c}{He2-104$^b$}&&& \multicolumn{3}{c}{R Aqr$^c$}&&& \multicolumn{3}{c}{CH Cyg$^d$}\\ \hline
\  \Ts (K) & \multicolumn{3}{c}{26500}&&& \multicolumn{3}{c}{130000}&&& \multicolumn{3}{c}{80000}&&& \multicolumn{3}{c}{150000-35000}\\
\  &rev & exp & lobe &&& rev& exp &  lobe &&& rev & exp & jets &&& rev & exp & jets\\ \hline
\   U        & 1  & 0 - 25  & 0. &&& 0.4   & 0.005& 0. & &&0.005-0.01 & - & 0.-0.0023 &&& 1-6 & 0.01-0.001&-\\
\  \Vs \, (\kms)& 400 &   70-190  & 210 &&& 300 & 50 & 250 &&& 110-120 & - & 50-150&&&600-1200&70-150& 70-100\\
\ \n0 (10$^4$\cm3) & 10   &0.16-20 & 2  && &7   & 2    &0.1&& & 6 & - & 0.005-7&&&1000-5000&10-1000 & 0.5 \\
\ \B0 (10$^{-3}$ g)& 1 & 0.1-1  & 1  & & & 1     & 1     & 1    & && 2     & - & 0.1-1 &&&3&1&3 \\
\ $D$ (10$^{15}$ cm)   &0.1 & 1-5          &100 &    &   & 4.5 &2. & $>$ 10 &&& 0.2 & - & 1.4-10&&&0.1 & 5-300&1  \\
\hline

\end{tabular}}

\flushleft

$^a$ this paper;

$^b$ Contini \& Formiggini (2001);

$^c$ Contini \& Formiggini (2003);

$^d$ results from Contini et al. (2009a).
\end{table*}

\section{Discussion and concluding remarks}

We  present  the quantitative results obtained  by the analysis of  the  spectra observed 
from BI Cru in different epochs.\\

\noindent
$\bullet$ {\bf 1962} :
we analysed the spectra observed in 1962  within a colliding-wind theoretical framework (Sect. 3).
 The data of the continuum were
observed many years later than those of the 1962 line spectra.
Even  considering a minimum set of prototypes,
the emerging picture of the BI Cru nebular network is consistent.

The result obtained by modelling the line spectra
indicates that in 1962 BI Cru may have been in a post-eruption "classical" phase,  with two main shock
networks created by the collision of the stellar winds.
The shock front between the stars, facing the hot star, has a velocity of \Vs=400 \kms, while the
expanding shock shows different components with \Vs= 70-190 \kms, in agreement with the observed FWHM.
The shock velocities are thus similar to those found in other SSs for the reverse and expanding shocks,
respectively (Angeloni et al. 2007a,b,c and references therein).

The physical conditions (e.g. n$_e$ $\geq$ 10$^6$ \cm3 downstream) which result from line modelling, are responsible
for self-absorption of free-free radiation in the radio range, and also indicate that both
permitted and forbidden FeII lines can be emitted from the same region.\\

\noindent
$\bullet$ {\bf1974} :
 we  argued that the observed broad blue wing of \Ha could not be emitted from the accretion disk
because the blue wing and the red wing show different polarization. So  we present in this paper
a new  interpretation of the \Ha broad blue wing adopting
the model  developed by Chevalier (1982) to explain the hydrodynamical picture in SNae after the explosion,
combined with the Heng \& Sunyaev (2008) theory of broad Lyman and Balmer emission line production.

In particular, the broad blue wing of the \Ha line observed by Whitelock et al. (1983) in 1974 is explained
by emission throughout the blast
wave shock  front created by a past ($\geq$60 years ago), unrecorded outburst which has now reached a 
radius  $\leq$ 1 pc.
Adopting  the Robinson et al. (1989) model of line formation
from the accretion disk,  the broad \Ha line emission should originate from the innermost regions
of the disk, being always perceived  only if the accretion disk is face-on, but hardly seen in
other configurations. On the other hand,
by  Chevalier's theory (Sect. 3.3)  the broad \Ha line is formed within the limit of the system
($\geq$  0.3 pc) and is seen only  from the side of the WD opposite to the Mira.

We  have demonstrated that,  even though no outbursts have  been recently observed in BI Cru
at least in the last 60 years of historical observations, this  does not invalidate our analysis;
on the contrary, this negative evidence is  exploited as a temporal constrain in the  calculations.

Interestingly, the  blast wave velocity ($\sim$ 5800-7700 \kms) estimated  for BI Cru at  
very early times after the outburst (Sect. 4),
is about twice the velocity of the nova ejecta in RS Ophiuci immediately after the 1985 outburst 
($\sim$ 2700-3900, Shore et al. 1996) and after the 2006 outburst
($\sim$ 3500 \kms, Walder et al. 2008). 
  However, in the broad-line phase of RS Ophiuci observed 6 days after the brightness peak,
 the OI 1300 and NIII] 1750 lines  showed   FWZI  of 7000 \kms (Shore et al. 1996).
Moreover, the detection of X-rays  from RS Ophiuci (e.g. O'Brien et al. 1992)
was explained by shocks. X-ray emission
could be expected also from BI Cru (Fig. \ref{fig:sed}).

Concluding, the spectrum observed from BI Cru in 1962 indicates that we were seeing the internal region 
between the stars.
The broad \Ha reported in 80's and 90's indicated that we were seeing the system from the side of
the WD opposite to the Mira.  We wonder whether these insights are sufficient to
suggest a likely
orbital period for BI Cru of  $\leq$ 100 years.

\noindent
$\bullet$ {\bf 1992}: 
 the lobes observed by SC92  in 1992, expanding out to 1.3 pc, are dominated by the [NII] 6584 line.
We have modelled the filaments in the lobes by constraining the model by the absence of [OII] 3727 and [OIII] 5007.
The fit of the calculated model to the observed continuum SED leads to a filling factor in the lobes of about 0.001.\\

Finally, in Table \ref{tab:t2} we compare the results obtained for BI Cru with those obtained for He2-104, R Aqr and CH Cyg, all objects showing lobes and jets extending outwards.

The temperatures of the hot star are within the range of moderately WD at quiescence
($\leq$ 30000 K - BI Cru) and those ($\geq$ 100000 K - CH Cyg)  that power the typical high ionization
emission line spectra of SSs; R Aqr, with \Ts=80000K, is just in between.
Ionization parameters, shock velocities, pre-shock densities, pre-shock magnetic fields, and geometrical thickness
of the disrupted filaments are rather similar in BI Cru and He2-104.
The relatively high geometrical thickness of the nebula downstream of the  reverse shock in He 2-104
would result in a large binary separation, but the large ranges of physical conditions in the expanding
fragments of BI Cru firmly indicate that its surrounding medium is actually more inhomogeneous.\\
The jets in R Aqr have a different aspect  due to a complex  combination of
shock dominated  and radiation dominated spectra with a low U,
 indicating strong dilution of the ionizing radiation.
By consistent modelling of the UV spectra emitted from the reverse nebula in R Aqr,
Contini \& Formiggini (2003)   confirmed that the inverse shock is
a standing shock and that a strong shock does not form in R Aqr, even if the lines show a
large FWHM.  In contrast, a strong shock
appears in BI Cru (\Vs=400 \kms), and  can be very strong (\Vs $\leq$ 1500 \kms)
in CH Cyg.

\section*{Acknowledgments}
The authors acknowledge many helpful conversations with Dina Prialnik. They also would like to thank the referee, R.L.M. Corradi, for helpful comments that improved the readability of the paper.

\label{lastpage}
\end{document}